\documentclass[conference]{IEEEtran}

\usepackage{amsmath,amssymb,amsfonts}
\usepackage{newtxtext,newtxmath}
\usepackage{graphicx}
\usepackage{booktabs}
\usepackage{algorithm}
\usepackage{algpseudocode}
\usepackage{tabularx}
\usepackage{array}
\usepackage{xcolor}
\usepackage{url}
\usepackage{cite}
\usepackage{xspace}
\usepackage[hidelinks]{hyperref}
\usepackage{needspace}
\usepackage{placeins}
\usepackage{float}
\usepackage{seqsplit}
\usepackage{microtype}
\graphicspath{{plots/}}

\makeatletter
\def\@IEEEabskeysecsize{\normalsize}
\long\def\@makecaption#1#2{%
\ifx\@captype\@IEEEtablestring%
  \normalsize\bgroup\par\centering\@IEEEtabletopskipstrut{\normalfont\normalsize #1}\\{\normalfont\normalsize #2}\par\addvspace{0.5\baselineskip}\egroup%
  \@IEEEtablecaptionsepspace
\else
  \@IEEEfigurecaptionsepspace
  \setbox\@tempboxa\hbox{\normalfont\normalsize {#1.}\nobreakspace\nobreakspace #2}%
  \ifdim \wd\@tempboxa >\hsize%
    \setbox\@tempboxa\hbox{\normalfont\normalsize {#1.}\nobreakspace\nobreakspace}%
    \parbox[t]{\hsize}{\normalfont\normalsize\noindent\unhbox\@tempboxa#2}%
  \else%
    \ifCLASSOPTIONconference \hbox to\hsize{\normalfont\normalsize\hfil\box\@tempboxa\hfil}%
    \else \hbox to\hsize{\normalfont\normalsize\box\@tempboxa\hfil}%
    \fi%
  \fi%
\fi}
\makeatother

\newcolumntype{Y}{>{\raggedright\arraybackslash}X}
\newcolumntype{L}[1]{>{\raggedright\arraybackslash}p{#1}}

\DeclareRobustCommand{\ProjectName}{\textnormal{AdaRare}}
\DeclareRobustCommand{\AdaRare}{\ProjectName\xspace}
\DeclareRobustCommand{\AFLpp}{\mbox{AFL++}\xspace}
\DeclareRobustCommand{\CmpLog}{CmpLog\xspace}
\DeclareRobustCommand{\LinUCB}{LinUCB\xspace}
\DeclareRobustCommand{\FuzzBench}{FuzzBench\xspace}
\DeclareRobustCommand{\DelegatedSharing}{clipped delegated-sample sharing\xspace}

\hypersetup{
  pdftitle={AdaRare: Telemetry-Guided Joint Profile Control for Greybox Fuzzing},
  pdfauthor={Jingchuan Ma, Tongan Liu, Yanhua Liu, Qiaoyun Huang},
  pdfsubject={Public preprint},
  pdfkeywords={greybox fuzzing, adaptive fuzzing, AFL++, telemetry-guided control}
}

\newcommand{\safeincludegraphics}[2][]{%
  \IfFileExists{#2}{%
    \includegraphics[#1]{#2}%
  }{%
    \fbox{%
      \begin{minipage}{0.92\linewidth}
        \centering
        \vspace{1.5em}
        Missing figure file: \texttt{\detokenize{#2}}.
        \vspace{1.5em}
      \end{minipage}%
    }%
  }%
}

\begin{document}

\title{\ProjectName: Telemetry-Guided Joint Profile Control for Greybox Fuzzing}
\author{%
\IEEEauthorblockN{Jingchuan Ma, Tongan Liu, Yanhua Liu, and Qiaoyun Huang}
\IEEEauthorblockA{%
Department of Computer Engineering, Fuzhou University Zhicheng College\\
Fuzhou, Fujian, China\\
\texttt{kiciot@qq.com}, \texttt{darkank@qq.com},\\
\texttt{Lyhwa@fzu.edu.cn}, \texttt{huangqy@fdzcxy.edu.cn}}
}
\IEEEspecialpapernotice{(Preprint)}
\maketitle

\begin{abstract}
Greybox fuzzers combine interacting queue, mutation, dictionary, energy, and comparison-solving control surfaces, while prior adaptive systems typically optimize other decision objects or control layers. We present \AdaRare, an \AFLpp extension that coordinates five internal actuation mechanisms as one bounded in-process profile updated every 5,000~ms. Completed-window, action-induced telemetry feeds an arm-local recency-weighted linear scorer and a profile-conditioned controller target. The scorer borrows the algebraic structure of disjoint \LinUCB, but serves as a closed-loop profile-ranking mechanism rather than a calibrated contextual-bandit action-value estimator or statistical confidence bound.

Across three sequential repeated-trial phases, Main provides broad integrated-system evidence: \AdaRare has higher median edge coverage than vanilla \AFLpp on all eight targets, with five Holm-significant comparisons. In the strongest matched result, Full \AdaRare has higher median edge coverage than CmpLog-matched \AFLpp on all five follow-up targets, with four Holm-significant comparisons. Batch A finds higher medians for telemetry-guided selection than fixed-context, random, and round-robin schedules in all 15 target--control comparisons, with 13 Holm-significant comparisons. The experiments do not establish independent No-A6-versus-Shadow or scarcity-bundle effects; A6 evidence is target-dependent and weakens under batch-wide correction. In an unmatched firmware case study, \AdaRare-generated inputs exposed five distinct memory-corruption findings, each reproduced in a separate environment and later assigned a CVE identifier. Controller-boundary compute P99 medians are below 6.5~ms for a five-second window; complete-boundary P99 medians including synchronous logging are below 14.7~ms. These measurements characterize boundary latency, not total system overhead.
\end{abstract}

\section{Introduction}
Coverage-guided greybox fuzzing is a fixed-budget security-testing process. \AFLpp combines queue management, power schedules, dictionaries, havoc mutations, favored-seed logic, and comparison solving in one high-throughput execution substrate~\cite{aflpp,fuzzing-survey}. These mechanisms interact: comparison solving consumes execution budget, queue bias changes which program regions receive attention, and dictionary-heavy mutations can be especially valuable under particular parser states. Under a fixed CPU budget, these control choices determine how much potentially security-relevant program behavior a campaign can explore.

Runtime adaptation is not new. AFLFast and EcoFuzz adapt seed energy~\cite{aflfast,ecofuzz}; MOpt and DARWIN adapt mutation distributions~\cite{mopt,darwin}; SLOPT applies bandit optimization to mutation schemes inside \AFLpp~\cite{slopt}; T-Scheduler adapts seed selection~\cite{tscheduler}; MobFuzz jointly selects objective combinations and power schedules~\cite{mobfuzz}; SYZVEGAS coordinates task and seed selection~\cite{syzvegas}; SEAMFuzz learns seed-cluster-specific mutation distributions~\cite{seamfuzz}; autofz and KRAKEN allocate parallel resources~\cite{autofz,kraken}; and xFUZZ composes input- and mutation-scheduling plugins~\cite{xfuzz}. These systems typically optimize a local policy, a plugin combination, or an external resource-allocation layer. \AdaRare instead studies a systems question: can multiple interacting \AFLpp control surfaces be coordinated as one bounded, low-frequency, in-process control object without replacing the underlying execution substrate?

\AdaRare divides a campaign into 5,000~ms windows. Each effective profile jointly sets dictionary probability, havoc intensity, favored-entry preference, new-seed preference, and scarcity-conditioned energy shaping. Completed-window telemetry is summarized into six dimensions and consumed by an arm-local recency-weighted linear scorer; a bounded profile-conditioned controller target updates the selected meta-arm and, for A6, the delegated effective profile. Five base profiles encode interpretable campaign postures, while A6 is a delegated portfolio meta-arm.

The telemetry vector $x_t$ is generated under the current selected/effective profile and guides the next window. The linear scores are operational ranking quantities rather than calibrated counterfactual action values, and the paper does not propose a new bandit algorithm. Effectiveness is assessed through repeated system-level comparisons and diagnostic scheduling controls.

\textbf{Contributions.} This paper makes three contributions.
\begin{itemize}
  \item \textbf{Joint in-process actuation of multiple \AFLpp control surfaces.} Every five seconds, one bounded profile coordinates queue preferences, dictionary/mutation intensity, and scarcity-conditioned energy shaping within the preserved \AFLpp substrate.
  \item \textbf{An audit-logged telemetry-guided controller with a source-level specification.} It distinguishes selected meta-arms from effective profiles, uses arm-local recency-weighted scoring, and logs boundary updates including A6 delegation.
  \Needspace{8\baselineskip}
  \item \textbf{Matched system evidence, diagnostic controls, explicit negative results, and an unmatched deployment case study.} The strongest controlled comparison evaluates Full \AdaRare against a CmpLog-matched substrate; separate profile-control experiments diagnose dynamic selection against simple schedules. The paper reports unresolved No-A6-versus-Shadow and scarcity-bundle effects, batch-wide multiplicity sensitivity, and five responsibly disclosed findings that later received CVE identifiers, without claiming comparative vulnerability yield.
\end{itemize}
\section{Related Work and Novelty Boundary}
\label{sec:related-positioning}

Prior fuzzers adapt different decision objects. Seed- and energy-oriented systems include AFLFast, Entropic, EcoFuzz, and T-Scheduler~\cite{aflfast,entropic,ecofuzz,tscheduler}. MOpt and DARWIN optimize mutation-operator probabilities~\cite{mopt,darwin}; SLOPT brings bandit optimization into \AFLpp at the mutation-scheme layer~\cite{slopt}; MuoFuzz learns conditional mutator sequences to exploit operator-order interactions~\cite{muofuzz}; and SEAMFuzz conditions mutation distributions on seed clusters~\cite{seamfuzz}. MobFuzz jointly adapts objective combinations and seed energy~\cite{mobfuzz}, and SYZVEGAS jointly adapts kernel-fuzzing task and seed selection~\cite{syzvegas}. Accordingly, adaptive fuzzing, bandit use in \AFLpp, and multi-control adaptation are not novel in isolation.

FOX formulates coverage-guided fuzzing as end-to-end online stochastic control and couples frontier-branch scheduling with a branch-distance-guided custom mutator and modified instrumentation~\cite{fox}. \AdaRare does not introduce branch-distance instrumentation or a replacement mutator; it uses completed-window telemetry to coordinate existing \AFLpp queue, dictionary/mutation, and energy actuation paths. We do not claim performance superiority over FOX without a direct comparison.

Composition and parallel systems operate at other layers. autofz assigns CPU time among complete fuzzers, and KRAKEN adapts parallel-fuzzing strategies and resources~\cite{autofz,kraken}. xFUZZ is the closest composition precedent: it selects runtime combinations of independently switchable input- and mutation-scheduling plugins~\cite{xfuzz}. TrioFuzz addresses adaptation latency architecturally by decoupling a centralized learner from parallel execution threads in a three-tier design, reducing its adaptation cycle to under one minute~\cite{triofuzz}. \AdaRare instead retains control inside one \AFLpp process and applies a compact joint internal profile every five seconds. We treat TrioFuzz as an architectural precedent for low-latency adaptation and do not claim superiority without a direct comparison.

\begin{table*}[t]
\caption{Decision-object taxonomy for representative adaptive fuzzers. The distinctions are architectural, not a performance ranking.}
\label{tab:related-taxonomy}
\centering
\renewcommand{\arraystretch}{1.04}
\begin{tabularx}{\textwidth}{@{}L{0.16\textwidth} L{0.29\textwidth} L{0.20\textwidth} Y@{}}
\toprule
Work & Decision object & Adaptation layer & Relation to \AdaRare \\
\midrule
AFLFast/EcoFuzz/T-Scheduler~\cite{aflfast,ecofuzz,tscheduler} & seed selection or energy & queue or power scheduling & adapts a local queue/power surface \\
MOpt/DARWIN/\allowbreak SLOPT/\allowbreak MuoFuzz~\cite{mopt,darwin,slopt,muofuzz} & mutation operators, schemes, or sequences & online mutation policy & mutation-local adaptation; MuoFuzz models sequence interactions \\
SEAMFuzz~\cite{seamfuzz} & seed-cluster mutation distribution & cluster-conditioned mutation & finer seed-conditioned mutation policy \\
MobFuzz~\cite{mobfuzz} & objective combination and power schedule & periodic multi-objective control & joint adaptation at a different objective layer \\
SYZVEGAS~\cite{syzvegas} & task and seed selection & online kernel-fuzzing policy & joint task/seed decision object \\
autofz/KRAKEN~\cite{autofz,kraken} & fuzzer/worker strategy and resources & parallel orchestration & allocates effort outside one execution substrate \\
xFUZZ~\cite{xfuzz} & input/mutation plugin composition & sliding-window composition & closest strategy-composition precedent \\
\AdaRare & joint profile over five \AFLpp-internal actuation mechanisms & fixed 5-s in-process control & coordinates queue, mutation/dictionary, and energy surfaces \\
\bottomrule
\end{tabularx}
\end{table*}

\AdaRare's technical boundary is joint in-process actuation of five \AFLpp-internal actuation mechanisms through one low-frequency closed-loop profile. Neither adaptive fuzzing, bandit algebra, nor runtime composition is novel in isolation; the linear scorer is an implementation mechanism rather than the paper's algorithmic contribution. The systems contribution is the realization, observability, and controlled evaluation of a profile spanning queue-selection preferences, mutation/dictionary intensity, and energy shaping without replacing the execution substrate. Learning-guided fuzzers use richer models and feedback sources~\cite{drlfuzz,datacoverage,shapfuzz,pathdiversity,predictive-context}; \AdaRare instead emphasizes a compact auditable controller and evaluates external edge coverage rather than internal scoring quantities. Because fuzzing is stochastic and implementation details matter~\cite{fuzzbench,fuzzing-reflections}, all inferential claims remain within target and batch.

\section{Background and Control Problem}

\subsection{Coverage-Guided Greybox Fuzzing}

A coverage-guided greybox fuzzing (CGF) system instruments a target, observes feedback such as edge coverage, retains interesting inputs, and mutates the queue to explore new behavior. A typical \AFLpp loop selects a queue entry, mutates it, executes the target, observes coverage and faults, and decides whether to save the input. Seed scheduling, power scheduling, mutation scheduling, dictionaries, and comparison solving all influence this loop. \AdaRare does not replace them; it adds an outer controller that biases existing mechanisms according to recent campaign telemetry.

\subsection{Joint Profile-Control Layer}

\AdaRare controls a scheduling profile rather than a seed, energy value, or mutation operator. This coarser layer is deliberate: AdaRare keeps controller updates outside the per-execution hot path and uses window-level telemetry to summarize the recent campaign regime beyond a single seed. This is a design choice rather than a claim that finer-grained online adaptation is infeasible. The window summary can capture whether coverage is still increasing, scarcity-weighted activity is changing, favored seeds dominate, or timeouts are rising, and can then select the next meta-arm and resolve its effective profile.

\subsection{Closed-Loop Profile-Control Formulation}

Let a fuzzing campaign have wall-clock budget $T$, divided into windows $t=1,2,\ldots$. Let $a_t\in\mathcal{A}=\{\mathrm{A1},\ldots,\mathrm{A6}\}$ denote the selected meta-arm, and let $e_t\in\mathcal{E}=\{\mathrm{A1},\ldots,\mathrm{A5}\}$ denote the effective base profile applied to \AFLpp during window $t$. If $a_t\neq\mathrm{A6}$, then $e_t=a_t$; when $a_t=\mathrm{A6}$, $e_t$ is sampled from A6's delegated top-$k$ set. Let $x_t\in\mathbb{R}^6$ summarize telemetry from the completed window, and let $r_t$ denote the bounded profile-conditioned controller target called ``reward'' in the source and audit logs. At the boundary, \AdaRare attributes $(x_t,r_t)$ to the selected decision, then uses $x_t$ to rank $a_{t+1}$ and, if needed, resolve delegated profile $e_{t+1}$.

This is a closed-loop profile-control problem: the selected meta-arm and effective profile directly affect the telemetry observed later. Dictionary-heavy behavior changes comparison activity, queue preferences change queue composition, and profile choices affect throughput and scarcity-weighted activity. The controller therefore uses completed-window, action-induced telemetry as a regime summary for the next decision. Its internal scores are not interpreted as unbiased counterfactual rewards under exogenous pre-action contexts; the external objective remains fixed-budget coverage growth and repeated-trial final edge coverage.

\subsection{Design Requirements}

A practical \AFLpp scheduler must adapt to non-stationary campaign phases, keep learning outside the target-execution hot path, prevent auxiliary proxy terms from dominating the bounded observable signals, and remain auditable through deterministic configuration, window logs, and numerical guardrails.

\section{\AdaRare Overview}

\subsection{Workflow}

\begin{figure*}[t]
\centering
\safeincludegraphics[width=0.98\textwidth]{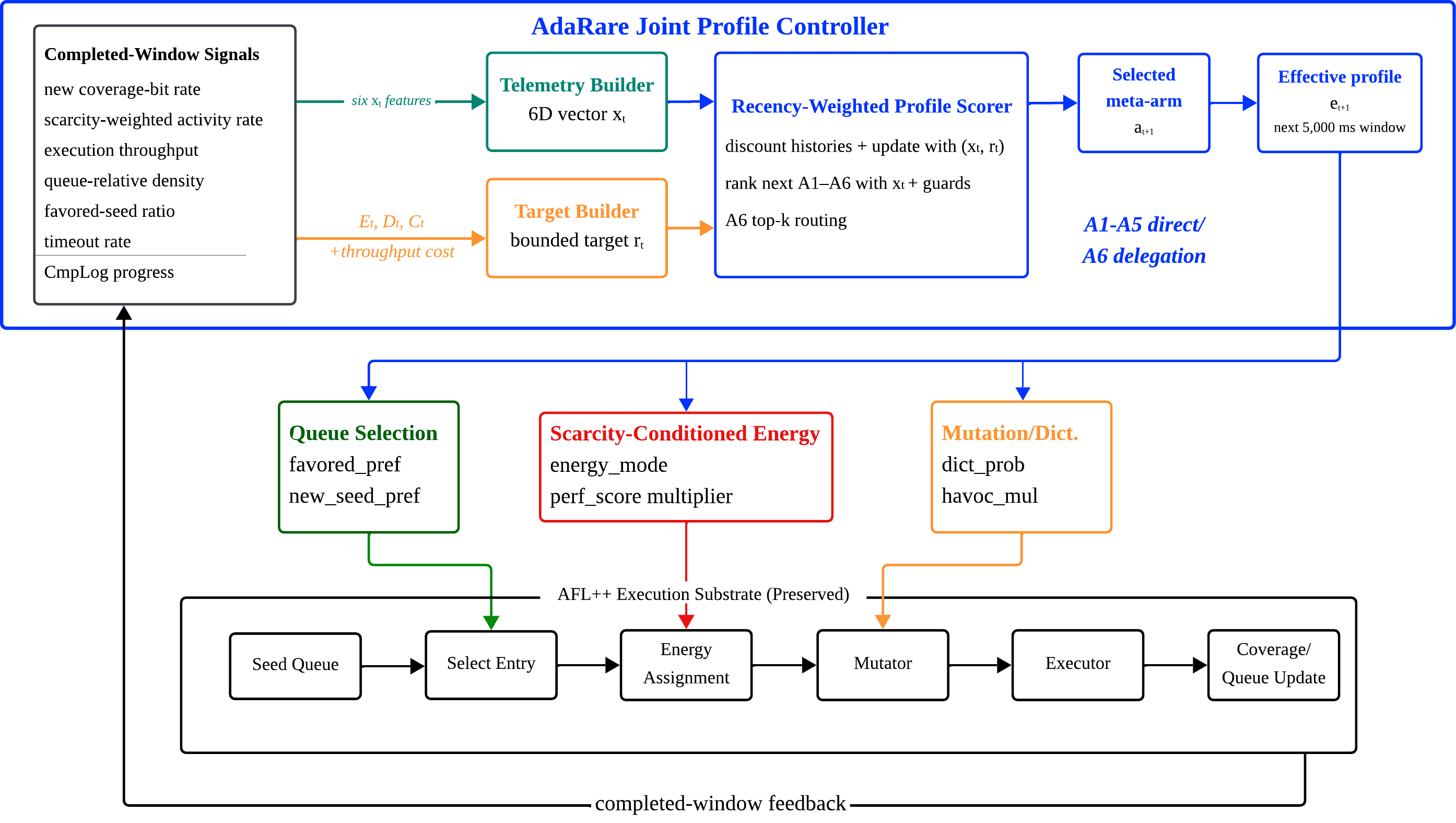}
\caption{\AdaRare as a joint profile controller over the preserved \AFLpp execution substrate. Completed-window telemetry forms $x_t$ and a bounded profile-conditioned controller target $r_t$. The profile scorer selects $a_{t+1}$ and resolves $e_{t+1}$, directly for A1--A5 or by delegation for A6. The effective profile jointly controls queue selection, mutation/dictionary behavior, and scarcity-conditioned energy shaping for the next window.}
\label{fig:overview}
\end{figure*}

\AdaRare is implemented as a window-level joint-control layer above the \AFLpp core loop. Figure~\ref{fig:overview} gives the system overview. During a window, \AFLpp performs its usual queue selection, mutation, execution, coverage collection, and queue update. \AdaRare accumulates window-level events. At the boundary, it constructs the telemetry vector, computes the profile-conditioned controller target, updates the arm-local scorer, selects the next meta-arm, resolves effective profile $e_{t+1}$, and writes that profile into the \AFLpp control paths.

\begin{figure*}[t]
\centering
\safeincludegraphics[width=0.98\textwidth]{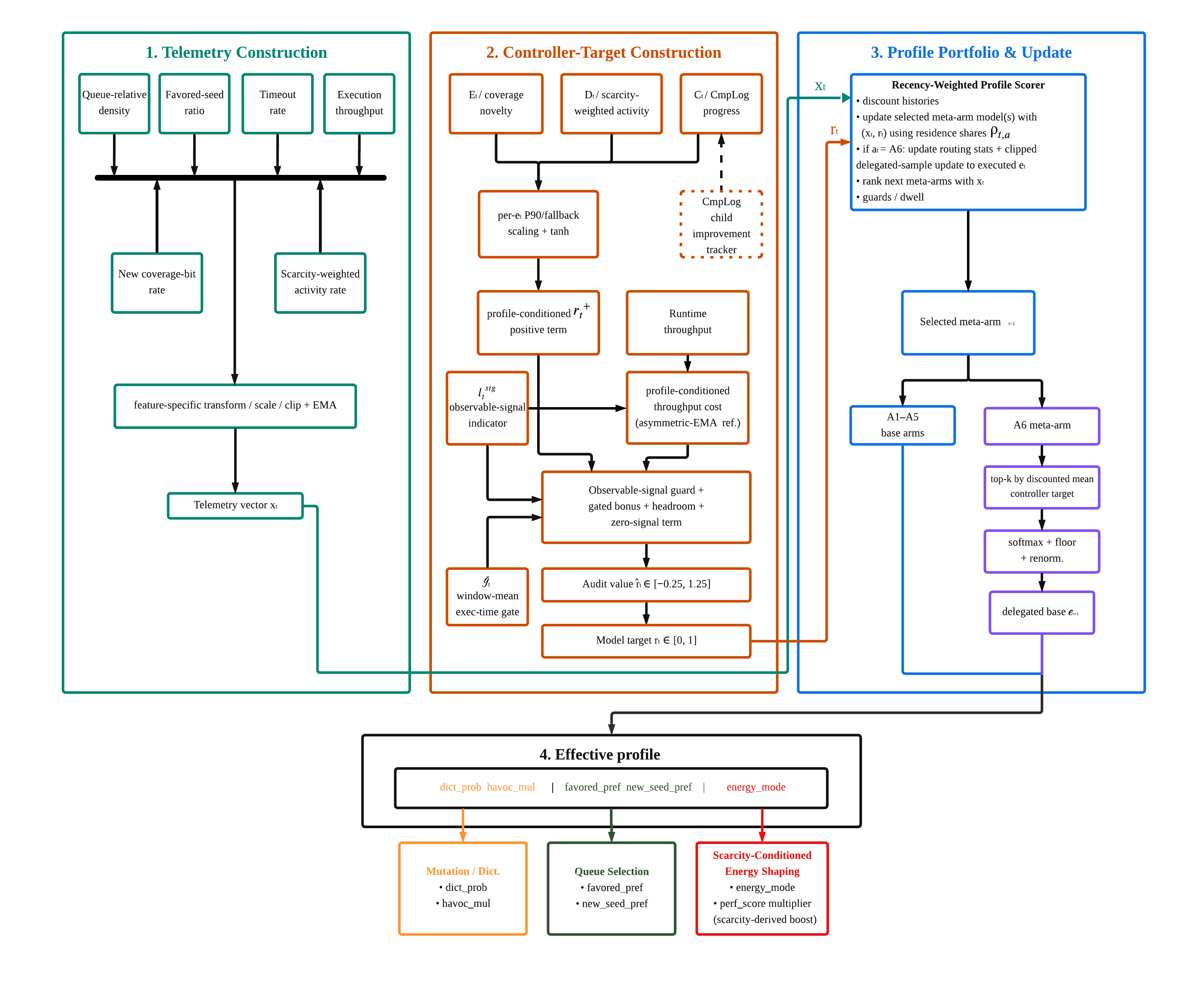}
\caption{Internal architecture of \AdaRare. Six runtime signals form action-induced telemetry $x_t$, while coverage novelty $E_t$, discovery-scarcity activity $D_t$, \CmpLog progress $C_t$, and throughput cost form profile-conditioned target $r_t$. The implementation's disjoint linear scorer selects meta-arm $a_{t+1}$ and resolves A1--A5 directly or A6 by delegation into effective profile $e_{t+1}$.}
\label{fig:internal-architecture}
\end{figure*}

The output of \AdaRare is not a concrete input or a replacement mutator. It is a profile that
changes the distribution of effort in the next window. Figure~\ref{fig:internal-architecture}
details this internal control path. This design keeps the integration local and makes the controller
easy to disable, audit, and compare against vanilla \AFLpp. It also enables profile-level
interpretation: the window logs expose when dictionary-heavy or queue-exploration profiles are selected.

\subsection{Telemetry Features}

At the end of each window, \AdaRare constructs a six-dimensional telemetry vector. The features are
coverage-bit discovery rate, discovery-scarcity-weighted activity rate, execution throughput, queue-relative density,
favored-seed ratio, and timeout rate. Raw feature values are log-compressed where appropriate,
scaled, capped, and smoothed with an exponential moving average:
\begin{equation}
  x_t = (1-\lambda_x)x_{t-1} + \lambda_x z_t,
\end{equation}
where $z_t$ is the raw feature vector for window $t$. The implementation uses $\lambda_x=0.30$.

The six features are chosen to cover the main feedback loops that a greybox campaign exposes without target-specific static analysis: novelty discovery, discovery-scarcity-weighted path activity, execution efficiency, queue pressure, \AFLpp favored-entry concentration, and target instability. This keeps the telemetry vector low-dimensional enough for auditable online updates while still representing campaign-level conditions that can trigger profile changes.

\begin{table*}[t]
\caption{Telemetry features used by \AdaRare. Features are scaled, capped, and smoothed before use by the profile scorer. Here, $Q_t$ is the queue size, $A_t$ is the number of active non-skipped queue entries, and $F_t$ is the number of entries marked favored by \AFLpp in window $t$.}
\label{tab:context}
\centering
\normalsize
\begin{tabularx}{\textwidth}{c l Y Y}
\toprule
Dim. & Feature & Definition & Intuition \\
\midrule
$x_0$ & New coverage-bit rate &
$\mathrm{clip}(\log(1+\Delta\mathrm{bits}_t/\tau_t)\cdot s_0,0,X_{\mathrm{cap}})$ &
Measures recent bitmap novelty. This is an internal signal, distinct from final \FuzzBench edge coverage. \\
$x_1$ & Scarcity-weighted activity rate &
$\mathrm{clip}(\log(1+D_t)\cdot s_1,0,X_{\mathrm{cap}})$ &
Measures path activity in slots with low historical virgin-bucket discovery counts; it is neither execution-frequency rarity nor a novelty count. \\
$x_2$ & Execution throughput &
$\mathrm{clip}(\log(1+N^{\mathrm{exec}}_t/\tau_t)\cdot s_2,0,X_{\mathrm{cap}})$ &
Captures the efficiency of the current profile. \\
$x_3$ & Queue-relative density &
$\mathrm{clip}(\log(1+Q_t/\max(A_t,1))\cdot s_3,0,X_{\mathrm{cap}})$ &
Summarizes queue pressure relative to active entries. \\
$x_4$ & Favored-seed ratio &
$\mathrm{clip}((F_t/\max(Q_t,1))\cdot s_4,0,X_{\mathrm{cap}})$ &
Measures the fraction of queue entries marked favored by \AFLpp. \\
$x_5$ & Timeout rate &
$\mathrm{clip}(\log(1+N^{\mathrm{to}}_t/\max(N^{\mathrm{exec}}_t,1))\cdot s_5,0,X_{\mathrm{cap}})$ &
Captures instability or excessive execution cost. \\
\bottomrule
\end{tabularx}
\end{table*}

The features are collected through existing runtime telemetry. The timeout feature is not treated as inherently negative by the profile scorer; a slow window may still be valuable if it produces coverage novelty, scarcity-weighted activity, or comparison progress. The controller-target construction determines when throughput cost is penalized.

\subsection{Profile Arms}

\AdaRare uses five base profile arms and one portfolio arm. Each base profile specifies five bounded mechanisms: dictionary-attempt probability, havoc multiplier, favored-entry preference, new-seed preference, and scarcity-conditioned energy mode. A6 does not apply a fixed profile; it delegates to a selected base profile, which becomes the effective arm for the window.

\begin{table*}[t]
\caption{\AdaRare profile arms. A1--A5 are base profiles; A6 delegates all five mechanism values to a top-$k$ base profile. Havoc values are per-site factors.}
\label{tab:arms}
\centering
\setlength{\tabcolsep}{2.8pt}
\begin{tabularx}{\textwidth}{@{}c L{0.15\textwidth} c c c c L{0.11\textwidth} Y@{}}
\toprule
Arm & Profile & dict (\%) & havoc/site & favored & new & energy mode & Scheduling bias \\
\midrule
A1 & throughput-first havoc & 5 & 1.20 & 0 & 1 & linear-soft, cap 2 & broad, throughput-oriented exploration \\
A2 & dictionary-heavy & 45 & 1.05 & 0 & 0 & logarithmic, cap 5 & token injection for structured inputs \\
A3 & CmpLog synergy & 20 & 1.10 & 1 & 0 & linear-medium, cap 3 & comparison and favored-entry emphasis \\
A4 & favored exploitation & 12 & 1.18 & 1 & 0 & thresholded, cap 3 & concentrated favored-entry exploitation \\
A5 & queue exploration & 8 & 0.95 & -1 & 1 & flat, cap 1 & new and less-favored queue exploration \\
A6 & top-$k$ portfolio & delegated & delegated & delegated & delegated & delegated & stochastic delegation using routing statistics \\
\bottomrule
\end{tabularx}
\end{table*}

\begin{table*}[t]
\caption{Runtime semantics of the five profile-controlled mechanisms. They are \AFLpp-internal controls built on existing queue, mutation, dictionary, and power-scheduling mechanisms, not five pre-existing command-line knobs.}
\label{tab:control-surfaces}
\centering
\renewcommand{\arraystretch}{1.04}
\begin{tabularx}{\textwidth}{@{}l L{0.18\textwidth} L{0.19\textwidth} Y@{}}
\toprule
Mechanism & Neutral/base behavior & Profile values & Exact runtime effect \\
\midrule
Dictionary probability & 100 permits each native dictionary attempt & integer 0--100 & Accept the attempt when a uniform integer in $[0,99]$ is below the profile value; 0 disables the attempt. \\
Havoc factor & $1.0$ leaves the current \texttt{perf\_score} unchanged at each site & positive per-site factor & The same factor is applied at two sequential score-scaling sites. The first result is integer-converted; the second uses upward-rounded percentage arithmetic. Each site caps \texttt{perf\_score} at $100\!\times\!\texttt{HAVOC\_MAX\_MULT}=6400$. \\
Favored preference & 0 accepts the alias-sampled entry & $-1,0,1$ & Prefer non-favored, neutral, or favored entries through at most eight rejection retries; alias weights are unchanged. \\
New-seed preference & 0 accepts the alias-sampled entry & $0,1$ & When enabled, prefer entries with \texttt{!was\_fuzzed} or zero fuzz level. If favored and new preferences are both active, both predicates must hold until the eight-retry cap, after which the current valid candidate is accepted. \\
Scarcity-conditioned energy & native \texttt{perf\_score} before AdaRare shaping & arm-specific bounded mode & Normalize the queue entry's discovery-scarcity score by a sampled P90 and multiply \texttt{perf\_score} by the effective profile's bounded boost; A6 inherits the delegated base mode. \\
\bottomrule
\end{tabularx}
\end{table*}

Table~\ref{tab:defaults} summarizes the main formal defaults. The Full campaigns used the defaults specified by the frozen formal controller configuration labeled \texttt{FORMAL-R1}; controller-parameter environment overrides and per-target settings were not used.

\Needspace{7\baselineskip}
\texttt{RELEASE-R1} is the internal label for a packaged 15-file method-relevant source subset, not a public release or historical-image proof. Appendix~\ref{app:config} reports the evaluated controller configuration, and Appendix~\ref{app:evidence} separates source-verifiable defaults, author-confirmed protocol facts, and unavailable run linkage.

\begin{table*}[t]
\caption{Main \AdaRare defaults. Values are fixed unless explicitly varied by a named control variant.}
\label{tab:defaults}
\centering
\begingroup
\renewcommand{\arraystretch}{1.06}
\setlength{\tabcolsep}{4.0pt}
\begin{tabularx}{\textwidth}{@{}L{0.16\textwidth} L{0.43\textwidth} Y@{}}
\toprule
Component & Evaluated default & Role \\
\midrule
Window and telemetry & 5,000~ms; telemetry EMA $0.30$ & Coarse boundary control and smoothing outside the per-execution hot path. \\
Warmup and dwell & two pulls per arm; one-window dwell threshold with guarded bypasses & Limits cold start and immediate unguarded switching. \\
Linear profile scorer & $\alpha=0.6$; ridge $10$; discount $0.999$ & Arm-local association, bounded exploration, conditioning, and recency weighting. \\
Running P90 scaling & 1024-sample uniform reservoirs; post-warmup and 32 positive samples & Per-effective-profile engineering scales for coverage, scarcity activity, and \CmpLog; a current positive observation is inserted before its window's P90 is evaluated. \\
Controller target and signal guard & raw caps $[-0.25,1.25]$; model target $[0,1]$; zero-signal term $0.015$ & Bounds the audit value and yields a zero, never a negative, model target when positive terms are absent. \\
\CmpLog & minimum gain $0.25$; clip $8.0$; fallback scale $10$ & Bounded post-execution comparison-map advancement. \\
A6 & top-$k=2$; temperature $1.20$; pre-normalization floor fraction $0.05$; sharing clip $2.0$ & Bounded stochastic delegation and additional executed-base credit. \\
\bottomrule
\end{tabularx}
\endgroup
\end{table*}

A profile is intentionally coarser than an individual mutation. The small action space supports interpretation, allows one update per window, and reuses \AFLpp mechanisms instead of introducing a replacement mutation engine.

\section{Design and Implementation}
\label{sec:design}

\subsection{Windows, Warmup, and Selection Guards}

\AdaRare evaluates its controller when monotonic wall time closes a 5,000~ms window, configured by \texttt{AFL\_BANDIT\_WINDOW\_MS}.

\Needspace{4\baselineskip}
Before telemetry-guided scoring, each arm receives two warmup observations, and the least-observed eligible arm is preferred. Dwell is a one-window anti-switch guard, but warmup, revisit, trend, and zero-signal emergency paths may bypass it. We do not treat dwell as an independent mechanism.

Algorithm~\ref{alg:window-controller} gives the source-level ordering. The selected meta-arm $a_t$ and effective base profile $e_t$ are distinct when A6 delegates. Telemetry from window $t$ is attributed only after that window completes; the resulting vector ranks profiles for window $t+1$.

\begin{algorithm}[t]
\caption{Window-boundary controller update and next-profile selection}
\label{alg:window-controller}
\begin{algorithmic}[1]
\Require completed window $t$ under selected meta-arm $a_t$ and effective profile $e_t$
\State Return unless monotonic wall time closes window $t$
\State Freeze $(a_t,e_t)$ and derive completed-window telemetry
\State Index edge, scarcity-activity, and \CmpLog reservoirs by $e_t$
\State Insert the current positive rates by uniform reservoir sampling
\State Then evaluate per-$e_t$ running P90 post-warmup after 32 positive samples; otherwise use fixed scales
\State Compute bounded controller-target components
\State Apply the observable-signal guard, headroom cap, and raw zero-signal term
\State Clip the audit value to $[-0.25,1.25]$ and model target $r_t$ to $[0,1]$
\State Discount all arm histories
\State Build and EMA-smooth action-induced telemetry $x_t$
\State Update selected-meta-arm models using residence shares $\rho_{t,a}$
\If{$a_t=\mathrm{A6}$}
  \State Update A6 routing statistics
  \State Give $e_t$ the guarded clipped update $w_t$ of Eq.~\eqref{eq:a6-sharing}
\EndIf
\State Rank eligible next meta-arms and apply warmup/revisit fallbacks
\State Enforce dwell unless an implemented guard bypasses it
\State Select $a_{t+1}$ and, for A6, sample $e_{t+1}$ from its top-$k$ set ($k=2$ here)
\State Emit the completed-window audit record
\State Save $(a_{t+1},e_{t+1})$ and apply queue/mutation/energy controls
\State Reset window accumulators
\end{algorithmic}
\end{algorithm}
\subsection{Recency-Weighted Telemetry-Guided Profile Scoring}

For each selected meta-arm $a$, \AdaRare maintains an arm-local design matrix $A_a$ and response vector $b_a$. The fitted parameter vector is
\begin{equation}
  \theta_a = A_a^{-1}b_a.
\end{equation}
Each arm is initialized with $A_a=\lambda_{\mathrm{ridge}}I$ and $b_a=0$. Given completed-window telemetry $x_t$, the implementation borrows the algebraic structure of disjoint \LinUCB~\cite{linucb}:
\begin{equation}
  \mathrm{score}_{t+1}(a\mid x_t)
  = x_t^\top\theta_a + \alpha\sqrt{x_t^\top A_a^{-1}x_t}.
\end{equation}
The first term is an arm-local fitted association between action-induced telemetry and the profile-conditioned controller target. The second is a bounded exploration term used to avoid premature concentration. Because $x_t$ is observed after the current profile acts, and because target scales and preferences depend on the effective profile, these scores are not calibrated counterfactual action values; the exploration term is not interpreted as a statistical confidence bound. The scores only rank eligible profiles inside the implemented closed-loop controller. All external effectiveness claims rely on repeated system-level comparisons.

A disjoint model preserves profile-specific response patterns across throughput-first, dictionary-heavy, comparison-oriented, queue-exploration, and portfolio behavior. To limit stale history, \AdaRare discounts the arm-local statistics at each window, following the recency-weighting idea used in non-stationary bandits~\cite{switching-ucb,switching-ts}. Let $\gamma_d\in(0,1]$ be the discount factor and $\lambda_{\mathrm{ridge}}$ the ridge term:
\begin{align}
  A_a &\leftarrow \gamma_d A_a + (1-\gamma_d)\lambda_{\mathrm{ridge}}I,\\
  b_a &\leftarrow \gamma_d b_a.
\end{align}
\Needspace{7\baselineskip}
Let $\rho_{t,a}$ be the fraction of recorded window residence time assigned to selected meta-arm $a$, with $\sum_a\rho_{t,a}=1$ in a complete window. Bookkeeping is normally one-hot. A recorded within-window transition is attributed through partial residence shares. These shares describe selected-meta-arm residence, not an effective-profile mixture.

The selected-arm models are updated as follows:
\begin{align}
  A_a &\leftarrow A_a + \rho_{t,a}x_tx_t^\top,\\
  b_a &\leftarrow b_a + \rho_{t,a}r_tx_t.
\end{align}
For the normal full-window case, $\rho_{t,a_t}=1$. If $a_t=\mathrm{A6}$, A6 receives this ordinary selected-meta-arm update, while the delegated effective profile $e_t$ receives the additional clipped update described below.

The implementation includes numerical guardrails. If matrix inversion fails, \AdaRare falls back to a non-contextual UCB-style ranking rule~\cite{ucb}. Exploration terms and final scores are bounded; matrix/vector values are rescaled if they grow too large; and finite-value checks prevent NaN or infinity propagation. These are implementation safeguards rather than independent mechanism claims.

\subsection{Profile-Conditioned Controller Target and Scarcity Activity}\label{sec:reward}

\noindent\emph{Discovery-scarcity-weighted activity.} For execution $i$, let $S_i$ be the nonzero classified bitmap-byte slots.

\Needspace{8\baselineskip}
Let $H_s$ be the campaign-lifetime count of previously virgin buckets/bits discovered at slot $s$. The count $H_s$ changes only when a new bucket/bit is exposed; it is neither an execution-hit count nor a path frequency. The evaluated configuration uses the following path-length-normalized mass:
\begin{equation}
 M_i=g_i\frac{1}{\max(|S_i|,1)}\sum_{s\in S_i}\frac{1}{\sqrt{H_s+1}}.
 \label{eq:scarcity-mass}
\end{equation}
Here, $g_i$ is the bounded execution-time gate defined below. Every nonempty execution can therefore contribute positive mass even without new coverage.

\Needspace{6\baselineskip}
Let $\tau_t$ denote the completed-window duration in seconds. The completed-window activity rate is
\begin{equation}
 D_t=\frac{\sum_{i\in t} M_i}{\tau_t}.
 \label{eq:activity-rate}
\end{equation}
This rate is a discovery-scarcity-weighted \emph{activity} signal, not novelty progress and not standard low-execution-frequency rarity. Window accumulation resets at each boundary; the $H_s$ table persists without decay.

The same normalized per-execution observation also updates the current queue entry's bounded scarcity score. At energy assignment, that score is decayed by $0.995$, capped at $10^9$, normalized by a sampled P90, and mapped through the effective profile's bounded energy mode from Table~\ref{tab:arms}. This fifth profile mechanism is distinct from the window-level controller target. \emph{No-scarcity-bundle} (display label: \emph{No-scarcity}; historical runtime label: \emph{No-rarity}) disables the combined scarcity telemetry feature, controller-target term, signal-guard input, and energy-shaping bundle.

For source and log correspondence, we retain the implementation notation $r_t$ and the term ``reward,'' but interpret it as a profile-conditioned controller target rather than a shared cross-profile utility. Its three nonnegative inputs are
\begin{equation}
 E_t=\frac{\Delta\mathrm{bits}_t}{\tau_t},\qquad
 D_t=\frac{\sum_{i\in t}M_i}{\tau_t},\qquad
 C_t=\frac{\Delta\mathrm{cmplog}_t}{\tau_t}.
\end{equation}
$E_t$ measures coverage-bit novelty, $D_t$ the scarcity-weighted activity above, and $C_t$ post-execution comparison-map progress. For the effective base profile $e_t$, separate size-1024 reservoirs maintain uniform samples of positive observations seen so far~\cite{reservoir}. A positive current-window rate is inserted before P90 evaluation and can therefore affect the scale used for the same window.

\Needspace{8\baselineskip}
After warmup and 32 positive samples, the running P90 replaces fallback scales $50$, $5$, and $10$ for $E_t$, $D_t$, and $C_t$; otherwise the fallbacks apply. The reservoirs are not rolling, decayed, or phase-local. Define
\begin{align}
 \nu_e&=\tanh(E_t/S_{e,e_t}),\\
 \nu_s&=\tanh(D_t/S_{s,e_t}),\\
 \nu_c&=\tanh(C_t/S_{c,e_t}).
\end{align}
These are per-effective-profile engineering scales, not cross-profile calibration factors.

\Needspace{6\baselineskip}
Controller-target preference weights are indexed by $e_t$. If A6 delegates to A3, both scales and weights come from A3; the resulting target updates selected A6 and gives delegated A3 an additional clipped update. A6 uses no separate preference row. Table~\ref{tab:reward-prefs} lists the five effective-profile rows.

\begin{table}[t]
\caption{Raw controller-target preferences of the effective base profiles. Each row is normalized by its nonnegative sum before use.}
\label{tab:reward-prefs}
\centering
\setlength{\tabcolsep}{4pt}
\begin{tabular}{c c c c c}
\toprule
Profile & $\omega_e$ & $\omega_s$ & $\omega_p$ & $\omega_c$ \\
\midrule
A1 & 0.55 & 0.15 & 0.30 & 0.35 \\
A2 & 0.35 & 0.50 & 0.15 & 0.35 \\
A3 & 0.45 & 0.40 & 0.15 & 0.70 \\
A4 & 0.70 & 0.20 & 0.10 & 0.35 \\
A5 & 0.40 & 0.35 & 0.25 & 0.35 \\
\bottomrule
\end{tabular}
\end{table}

\Needspace{12\baselineskip}
After row normalization, the positive controller-target term and throughput cost are
\begin{align}
 r_t^+ &= \tilde\omega_e\nu_e+\tilde\omega_s\nu_s+\tilde\omega_c\nu_c,\\
 \mathrm{cost}_t &= \tilde\omega_p P_t^{\mathrm{eff}},\\
 P_t&=\max\!\left(0,1-\frac{\mathrm{thrpt}_t}{\mathrm{thrpt}_{\mathrm{ref}}}\right),\qquad \mathrm{thrpt}_{\mathrm{ref}}>0.
\end{align}
The implementation sets $P_t=0$ if the reference is nonpositive. The throughput reference uses asymmetric updates.

The current-sample weight is $0.1$ when throughput rises (and during warmup) and $0.001$ when it falls. Let $I_t^{\mathrm{sig}}$ denote the presence of any configured positive observable signal: new coverage, new coverage bits, positive scarcity activity, or positive \CmpLog progress. Because $D_t$ may be positive without novelty, this is deliberately a \emph{signal indicator}, not a progress or novelty indicator.

\Needspace{5\baselineskip}
The effective throughput cost is
\begin{equation}
 P_t^{\mathrm{eff}}=P_t\begin{cases}0.10,&I_t^{\mathrm{sig}}=1,\\1,&I_t^{\mathrm{sig}}=0.\end{cases}
 \label{eq:throughput-cost}
\end{equation}
Every execution with a nonempty classified bitmap has $M_i>0$.

\Needspace{3\baselineskip}
Consequently, an active window containing such an execution normally has $D_t>0$ and hence $I_t^{\mathrm{sig}}=1$. The indicator therefore primarily distinguishes inactive or telemetry-empty windows; it is not a coverage-progress or novelty gate.

Equation~\eqref{eq:time-gate} defines the evaluated execution-time gate:
\begin{equation}
 g_i=\mathrm{clip}\!\left(\frac{1}{1+\eta_g\max(\mathrm{exec}_{i}/\mathrm{baseline}_{i}-1,0)},g_{\min},1\right).
 \label{eq:time-gate}
\end{equation}
Here, $\eta_g=1$ and $g_{\min}=0.05$. The execution-time baseline is initialized from the first measured execution and, before the corresponding gate evaluation, updated on later executions as $B_i=0.99B_{i-1}+0.01\,\mathrm{exec}_i$; the gate uses $\mathrm{baseline}_i=B_i$. Let $\bar g_t$ be the window mean and
\begin{equation}
 b_t=\mathrm{clip}\!\left(0.02\log(1+\bar g_t),0,0.15\right).
\end{equation}
With positive-target cap $R^+_{\max}=1.25$, define the remaining positive headroom
\begin{equation}
 h_t=\max\!\left(R^+_{\max}-(r_t^+-\mathrm{cost}_t),0\right).
\end{equation}
The effective multiplicative bonus is
\begin{equation}
 g_t^{\mathrm{eff}}=\begin{cases}
 \min\!\left(b_t,h_t/r_t^+\right), & I_t^{\mathrm{sig}}=1\ \text{and}\ r_t^+>10^{-4},\\
 0, & \text{otherwise.}
 \end{cases}
\end{equation}
Under the evaluated defaults, $g_i\leq 1$ implies $\bar g_t\leq 1$, so $b_t\leq0.02\log 2\approx0.0139$. Row normalization also bounds the positive term by its largest profile-wise positive-weight share, attained by A4: $r_t^+\leq(0.70+0.20+0.35)/(0.70+0.20+0.10+0.35)=25/27\approx0.9259$. Since $\mathrm{cost}_t\geq0$, whenever $r_t^+>0$ we have $h_t/r_t^+\geq0.35$. Thus neither the configured $1.15$ hard factor cap nor the positive-headroom cap is binding under the evaluated defaults; before subsequent raw/model clipping the multiplicative bonus is at most $1+0.02\log 2\approx1.0139$.
The raw audit value and model target are
\begin{align}
 r_t^{\mathrm{raw}}&=r_t^+(1+g_t^{\mathrm{eff}})-\mathrm{cost}_t-(1-I_t^{\mathrm{sig}})P_{\mathrm{zs}},\\
 \hat r_t&=\min(\max(r_t^{\mathrm{raw}},-0.25),1.25),\\
 r_t&=\min(\max(\hat r_t,0),1),
 \label{eq:model-target}
\end{align}
with $P_{\mathrm{zs}}=0.015$. When no positive signal is present, the raw value is negative but the model target is floored to zero. The zero-signal term therefore affects audit telemetry and guard state; it does not inject a negative linear-model target.

\subsection{Closed-Loop Signal Semantics}

Several \AdaRare signals are action-induced. For example, a dictionary-heavy profile can change
comparison activity, a favored-seed profile can change queue composition, and a slow profile can
affect both throughput telemetry and throughput penalty. This is intentional: the controller observes completed-window telemetry after executing an effective profile and then selects the next meta-arm and resolves its effective profile. It also means that
internal scoring signals must not be interpreted as external outcomes.

\AdaRare constrains this feedback loop in three ways. First, external evaluation uses repeated-trial \FuzzBench edge coverage rather than the internal controller target. Second, bounded scaling, clipping, and finite-value checks constrain each term.

\Needspace{6\baselineskip}
Third, window logs expose bounded target components and selected/effective profiles. Scarcity-weighted activity can itself create a positive auxiliary term without new coverage; we therefore do not describe the guard as proving progress or preventing all activity-only reward. The controlled campaigns, not the internal coefficients or controller target, support the external effectiveness claims.

\subsection{\CmpLog Progress and the A6 Portfolio Arm}
\label{sec:cmplog-a6}

\AdaRare uses \CmpLog as a post-execution progress signal. It does not predict which comparisons will be useful before mutation. The real \CmpLog child execution first fills \texttt{cmp\_map}.

\Needspace{5\baselineskip}
\AdaRare then examines observed comparison behavior and accumulates progress only when a comparison site improves over the best state seen in the current runtime tracker. This signal is derived solely from comparison-map advancement. Queue growth and ordinary coverage discovery are recorded separately and are not counted as \CmpLog progress.

For instruction comparisons, \AdaRare records reductions in best integer absolute distance when the
width is at most eight bytes and the comparison is not floating-point. It also records improvements
in equal-byte matches over up to 32 bytes. For comparison-function entries, it records equal-byte
improvements over up to 32 bytes. The raw comparison progress is compressed as
\begin{equation}
  \log(1+\Delta\mathrm{dist}_t)+0.5\log(1+\Delta\mathrm{match}_t),
\end{equation}
then filtered by a minimum-gain threshold and capped. This signal complements new coverage by adding controller-target mass when inputs move closer to satisfying structured comparisons before they unlock new edges.

A6 is designed to reduce dependence on any single base profile as campaign conditions change.
When A6 is selected, \AdaRare builds a top-$k$ set of base arms using discounted mean controller targets as routing statistics; the default is $k=2$. It samples from that set by softmax, applies a pre-normalization floor, and renormalizes before delegation. The delegated arm supplies the effective profile applied to \AFLpp during that window.

Both normalization reservoirs and controller-target preference weights are indexed by the delegated effective profile $e_t$. The resulting target is then used for A6's ordinary selected-meta-arm update. Let $j=e_t$ denote the delegated base arm. It receives an additional \emph{clipped delegated-sample} update:
\begin{equation}
  w_t=\min\left(\frac{1}{\max(\pi_t(j),10^{-6})},w_{\max}\right),
  \label{eq:a6-sharing}
\end{equation}
where $\pi_t(j)$ is the renormalized delegation probability, the denominator guard is $10^{-6}$, and $w_{\max}=2.0$.
The delegated arm then receives
\begin{align}
  A_j &\leftarrow A_j + w_t x_t x_t^\top,\\
  b_j &\leftarrow b_j + w_t r_t x_t.
\end{align}

This rule is narrower than a general importance-sampling estimator. It credits only the base profile that actually controlled the window; unexecuted profiles receive no synthetic controller target. The inverse-probability form is only a heuristic credit multiplier for unequal delegation probabilities; clipping bounds individual-sample influence. Neither the unclipped nor clipped form is interpreted as an unbiased off-policy estimator. We therefore call the mechanism \DelegatedSharing rather than a full off-policy estimator.

\subsection{Integration and Observability}

\AdaRare is a local \AFLpp extension. It receives coverage, discovery-scarcity activity, execution, timeout, slow-execution, and \CmpLog events, and writes profile parameters for dictionary probability, havoc intensity, favored/new-entry preference, and scarcity-conditioned energy shaping. Default profiles do not alter trimming.

The prototype is based on \AFLpp 4.35c. It exports \texttt{.adarare\_bandit.csv},
\texttt{.adarare\_overhead.csv}, \texttt{.adarare\_config.json}, and optional verify/debug logs.
The bandit CSV records source-named context and reward components, scaling values, selected arms, delegated A6
probabilities, throughput penalties, zero-signal state, and numerical fallback counters. The
overhead sidecar records window-boundary timing fields, including telemetry construction, controller-target
computation, profile scoring, model update, profile application, audit-log emission, campaign-level
execution rate, and selected/effective arm identifiers. The JSON file records static controller
configuration when archived. These outputs are not used for decision-making, but they are necessary
for reproducibility, ablation analysis, controller-activation checks, boundary-latency measurement, and
failure diagnosis.

\section{Evaluation}
\label{sec:evaluation}

We ask five controlled-evaluation questions: \textbf{RQ1} what broad integrated-system coverage evidence Main provides; \textbf{RQ2} whether that separation persists across repeated trials and time; \textbf{RQ3} how dynamic telemetry-guided selection compares with simple schedules within the fixed A1--A5 action space; \textbf{RQ4} whether the integrated result survives a matched \CmpLog substrate and what the mechanism diagnostics establish; and \textbf{RQ5} how large the controller-boundary latency is, with end-to-end execution rates reported only as context. We separately report an unmatched real-world deployment case study.

\subsection{Datasets, Provenance, and Statistics}
Table~\ref{tab:batches} separates three sequential evaluation phases. We first ran the eight-target Main study, then added the 500-campaign Batch A profile-control follow-up, and finally the 300-campaign Batch B fairness/mechanism follow-up. The integrated Full controller semantics and A1--A6 profiles were unchanged; intervening commits added experimental control modes, FuzzBench/environment wiring, and packaging. Each controlled batch fixed one revision across its compared variants. We nevertheless analyze every phase independently and never pool absolute edge counts across batches.

\begin{table*}[t]
\caption{Final experimental phases. All exports contain ten trials per cell and complete 900-second snapshots through 86,400 seconds; ``Primary endpoint'' identifies the endpoint used for each phase's inferential families.}
\label{tab:batches}
\centering
\begin{tabular}{lrrrrll}
\toprule
Phase & Targets & Variants used & Campaigns used & Duration & Primary endpoint & Purpose \\
\midrule
Main & 8 & 7 & 560 & 24 h & 86,400 s & broad integrated-system evidence \\
A & 5 & 10 & 500 & 24 h & 85,500 s & scheduling-policy diagnostics and A6 \\
B & 5 & 6 & 300 & 24 h & 85,500 s & strongest matched evidence and mechanism diagnostics \\
\bottomrule
\end{tabular}
\end{table*}

The Main export contains 800 campaigns. The aligned seven-fuzzer analysis uses 560 campaigns. The other 240 form three complete 80-campaign legacy matrices (No-Cmp, No-Context, and No-IPS).

\Needspace{6\baselineskip}
We exclude complete cells rather than individual trials because those historical variants lack the same-commit attribution controls used in the follow-ups.

\Needspace{8\baselineskip}
The five-target controlled subset was selected after Main. Its original selection rationale was not preserved in a pre-outcome protocol, so Batches A and B are controlled follow-up evidence rather than an outcome-blind replication. The subset contains three Main-significant targets (\texttt{curl}, \texttt{lcms}, and \texttt{openh264}) and two Main-nonsignificant targets (\texttt{proj4} and \texttt{sqlite3}); this does not remove selection-bias risk. It spans network/structured parsing, color processing, media decoding, geospatial processing, and database processing.

The Main study uses the complete 86,400-second endpoint.

\Needspace{9\baselineskip}
Batches A and B retain the 85,500-second inferential endpoint recorded in the preserved analysis protocol; their complete 86,400-second observations are sensitivity analyses in Appendix~\ref{app:endpoint}. We do not claim that 85,500 seconds is intrinsically preferable to a full day. Directions and within-family Holm-significance counts are unchanged. Secondary views use 6 hours, 12 hours, and normalized area under the coverage curve through the corresponding primary endpoint. Tests use SciPy~1.17.0 two-sided Mann--Whitney U with \texttt{method=auto}; at 10-versus-10, this is the asymptotic branch. Tie and default continuity corrections are applied. Holm correction is applied within each estimand-defined family; unless stated otherwise, significance means Holm-adjusted $p<0.05$ in the specified family. We interpret MWU as a rank/distribution comparison; median gains are descriptive rather than the tested null~\cite{evalfuzz,prudentfuzz}. We report Cliff's $\delta$ for the primary AdaRare/AFL++ comparison.

\Needspace{3\baselineskip}
The prepared artifact package retains Vargha--Delaney $\hat A_{12}$ for the other comparisons. Deterministic 10,000-resample \emph{independent percentile} bootstrap intervals are descriptive only. Appendix~\ref{app:holm} also applies one conservative Holm correction to all 45 Batch A or 35 Batch B contrasts. All ten trials in every analyzed cell are retained; none is deleted or imputed.

The Main phase used a rented cloud allocation reported as 128 CPUs/vCPUs, 256~GB RAM, and 2~TB storage; its CPU model was not retained. Batches A and B used a rented allocation reported as AMD EPYC~9754, 96 CPUs/vCPUs, 120~GB RAM, and 1.12~TB storage. These are author-retained allocation records rather than machine-verified topology. OS/kernel, pinning, and per-campaign worker-allocation metadata were not preserved.

Within Batch B, one canonical builder produced the target, \CmpLog binary, dictionaries, and effective time-zero corpus copied across all six variants for each target. CmpLog-matched enables the same \texttt{-c} substrate while disabling the controller. Shadow performs telemetry and controller-target construction, model updates, candidate selection, and logging but suppresses profile application. No-A6 keeps active telemetry-guided A1--A5 control and disables only the A6 portfolio. No-scarcity-bundle (displayed as No-scarcity; historical runtime label: No-rarity) is based on No-A6 and disables the scarcity telemetry feature, controller-target contribution, signal-guard contribution, and scarcity-conditioned energy adjustments while retaining coverage, \CmpLog, throughput, model update, selection, and profile application. These variant semantics are author-confirmed from the frozen experiment definitions and smoke/runtime checks; Appendix~\ref{app:evidence} records the retained evidence and missing run linkage.
\subsection{RQ1: Broad Integrated-System Coverage}
\begin{figure*}[t]
\centering
\safeincludegraphics[width=0.98\textwidth]{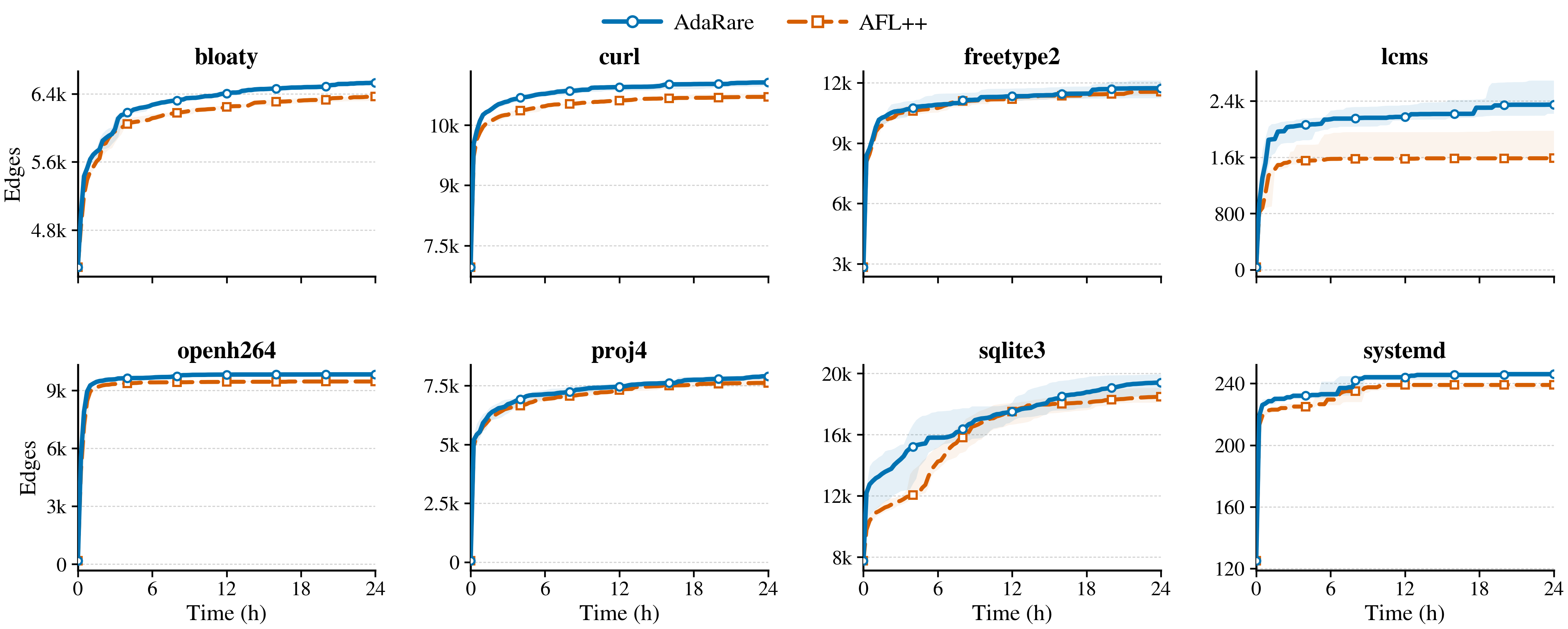}
\caption{Median edge coverage over 24 hours for the Main-phase integrated AdaRare/AFL++ comparison; bands show the interquartile range across ten trials. Color, line style, and marker shape redundantly encode each system for grayscale readability.}
\label{fig:coverage}
\end{figure*}

\begin{table*}[t]
\caption{Integrated AdaRare versus vanilla AFL++ at 86,400 seconds in the aligned 560-campaign, seven-scheduler subset of the 800-campaign Main export. Each cell reports median edge coverage with IQR width in brackets over ten trials. $p_H$ is Holm-adjusted across eight targets; $\delta$ is Cliff's delta.}
\label{tab:main-stats}
\centering\setlength{\tabcolsep}{3.2pt}
\begin{tabular}{lrrrrrr}
\toprule
Target & AdaRare & AFL++ & Gain & $p_H$ & $\delta$ & Result \\
\midrule
\texttt{bloaty} & 6531.5 [52.5] & 6367 [76] & +2.58\% & 0.0013 & 1.00 & sig. \\
\texttt{curl} & 11567 [137.5] & 11208 [118.2] & +3.20\% & 0.0023 & 0.92 & sig. \\
\texttt{freetype2} & 11729.5 [877.5] & 11552 [396.8] & +1.54\% & 0.6232 & 0.14 & n.s. \\
\texttt{lcms} & 2347.5 [472] & 1586 [396.8] & +48.01\% & 0.0016 & 0.96 & sig. \\
\texttt{openh264} & 9828 [23.2] & 9460.5 [89.8] & +3.88\% & 0.0013 & 1.00 & sig. \\
\texttt{proj4} & 7896.5 [226.8] & 7612 [331.5] & +3.74\% & 0.3637 & 0.42 & n.s. \\
\texttt{sqlite3} & 19398 [1351.5] & 18472 [958.2] & +5.01\% & 0.3637 & 0.42 & n.s. \\
\texttt{systemd} & 246 [1.8] & 239 [2.2] & +2.93\% & 9.5e-04 & 1.00 & sig. \\
\bottomrule
\end{tabular}
\end{table*}

At 86,400 seconds, the integrated \AdaRare configuration has a higher median than vanilla \AFLpp on all eight targets; five are Holm-significant. Gains range from 1.54\% on \texttt{freetype2} to 48.01\% on \texttt{lcms}. The three nonsignificant targets are \texttt{freetype2}, \texttt{proj4}, and \texttt{sqlite3}; we report them as positive but inconclusive. Because Main is not \CmpLog-matched, it is broad integrated-system evidence rather than scheduler-only attribution.

Heterogeneous scheduler results are provided in Appendix~\ref{app:heterogeneous} as context and are not used for a superiority claim. The strongest controlled claim is the matched Full/CmpLog comparison in RQ4; Main provides broader integrated-system context, and the prepared artifact package retains the full seven-scheduler time series, final-trial distributions, and per-comparison statistics.

\subsection{RQ2: Repeated-Trial Stability}
Figure~\ref{fig:coverage} shows that the separation is not confined to the final snapshot: the AdaRare median and IQR usually separate progressively from AFL++. Table~\ref{tab:main-stats} also shows compact IQRs on \texttt{bloaty}, \texttt{curl}, \texttt{openh264}, and \texttt{systemd}. \texttt{lcms} combines the largest median gain with a wider IQR than AFL++, so we describe it as a high-gain, higher-variance case. Stability is not equated with low variance alone; a consistently low-coverage scheduler can be stable but ineffective.

\subsection{RQ3: Diagnostic Profile-Selection Controls (Batch A)}
\label{sec:rq3-batch-a}
\begin{table}[t]
\caption{Batch A profile-control results at the frozen 85,500-second inferential endpoint. The first three rows are corrected jointly within the 15-test controller family; Full versus No-A6 uses the separate five-test A6 family.}
\label{tab:batch-a-summary}
\centering\setlength{\tabcolsep}{3.2pt}
\begin{tabular}{lrrl}
\toprule
Comparison & Higher & Holm sig. & Gain range \\
\midrule
No-A6 vs Constant & 5/5 & 4/5 & 4.84--8.73\% \\
No-A6 vs Random & 5/5 & 5/5 & 10.34--15.54\% \\
No-A6 vs Round-robin & 5/5 & 4/5 & 3.70--14.17\% \\
Full vs No-A6 & 5/5 & 2/5 & 1.37--4.56\% \\
\bottomrule
\end{tabular}
\end{table}

Batch A fixes A1--A5 and disables A6 in the controller controls. No-A6 retains the six-dimensional dynamic telemetry vector.

\Needspace{7\baselineskip}
Constant-context replaces dynamic telemetry with the fixed vector $(1,0,0,0,0,0)$ for both model updates and next-profile scoring. It otherwise preserves the arm-local linear models, controller-target and parameter updates, fitted-association and exploration terms, candidate selection, and A1--A5 profile application. Random and Round-robin replace the telemetry-guided guarded scorer.

\Needspace{3\baselineskip}
They use random choice or cyclic selection over A1--A5, respectively; the five Static variants hold one base profile for the campaign.

\Needspace{10\baselineskip}
All controls retain the five-second cadence, runtime telemetry, controller-target construction, model bookkeeping, common A1--A5 definitions, and profile application; each Static arm also retains its profile-specific scarcity-conditioned energy mode. At 85,500 seconds, No-A6 has a higher median than Constant-context, Random, and Round-robin on every target. Holm-significant counts are 4/5, 5/5, and 4/5, respectively.

\Needspace{4\baselineskip}
Together, 13/15 comparisons are Holm-significant in the joint controller family.

Against all five static arms, No-A6 has a higher median in 24/25 target--arm comparisons, with 21/25 Holm-significant comparisons. When the highest-median static arm is selected separately for each target, its median is lower than No-A6 on 4/5 targets; the exception is \texttt{lcms} with Static A2. Full has a higher median than No-A6 on all five targets, but the A6 increment is Holm-significant on two.

\Needspace{12\baselineskip}
Within the fixed A1--A5 action space, all 15 fixed-context, random, and round-robin target--control comparisons have higher medians under dynamic telemetry-guided selection, and 13 are Holm-significant. Most fixed-profile comparisons also favor dynamic selection. These are diagnostic results within the implemented action space, not evidence of superiority over alternative adaptive algorithms. A6 remains target-dependent in this batch.

\subsection{RQ4: Strongest Matched Evidence and Mechanism Diagnostics (Batch B)}
\label{sec:rq4-batch-b}
\begin{table*}[t]
\caption{Batch B estimand-defined comparison families at the frozen 85,500-second inferential endpoint. ``Higher'' counts targets with a higher treatment median; Holm correction is applied separately within the system-fairness (15 tests), substrate-diagnostic (10), scarcity-bundle (5), and A6 (5) families.}
\label{tab:batch-b-families}
\centering\setlength{\tabcolsep}{3.5pt}
\begin{tabular}{llrrl}
\toprule
Family & Comparison & Higher & Holm sig. & Median-gain range \\
\midrule
system fairness & Full vs AFL++ default & 4/5 & 4/5 & -0.07--+43.41\% \\
system fairness & Full vs CmpLog-matched & 5/5 & 4/5 & +0.54--+25.56\% \\
system fairness & No-A6 vs Shadow & 4/5 & 0/5 & -0.40--+19.79\% \\
substrate diagnostics & CmpLog-matched vs Default & 4/5 & 2/5 & -0.60--+30.76\% \\
substrate diagnostics & Shadow vs CmpLog-matched & 2/5 & 0/5 & -2.72--+0.63\% \\
scarcity bundle & No-A6 vs No-scarcity & 3/5 & 0/5 & -0.24--+17.59\% \\
A6 & Full vs No-A6 & 5/5 & 3/5 & +0.31--+10.82\% \\
\bottomrule
\end{tabular}
\end{table*}

Full has a higher median than default \AFLpp on four targets, each Holm-significant; \texttt{sqlite3} is near-equal and non-significant.

This is the paper's strongest matched system-level result: Full has higher median edge coverage than CmpLog-matched \AFLpp on all five targets, with four Holm-significant comparisons and median gains from 0.54\% to 25.56\%. Within Batch B's matched substrate, supplying \texttt{-c} and incurring the \CmpLog execution cost alone does not reproduce Full \AdaRare's coverage result.

The remaining mechanism results are more nuanced. No-A6 has a higher median than Shadow on four targets, but none of the five rows for this pair is significant after Holm correction within the 15-test system-fairness family. Shadow and CmpLog-matched are also close: only two targets have a higher Shadow median and none is significant. We therefore do not claim that controller computation alone helps, or that applying A1--A5 without A6 has an independently significant effect in Batch B.

No-A6 has a higher median than No-scarcity on only three targets, and none of the five comparisons is significant. This does not establish an independent scarcity-bundle effect; the bundle may interact with A6 or campaign phase, and the matrix lacks an A6-enabled/no-scarcity cell.

\Needspace{8\baselineskip}
The present five-target experiment does not isolate a stable independent coverage contribution from the combined scarcity telemetry feature, controller-target term, signal guard, and energy shaping. Those four subcomponents remain unresolved. In contrast, Full has a higher median than No-A6 on all five targets, with three Holm-significant comparisons and median gains of 0.31--10.82\%. Batch B therefore provides the clearest, but target-dependent, incremental evidence for A6 in this incomplete factorial matrix.

\begin{figure*}[t]
\centering
\safeincludegraphics[width=0.98\textwidth]{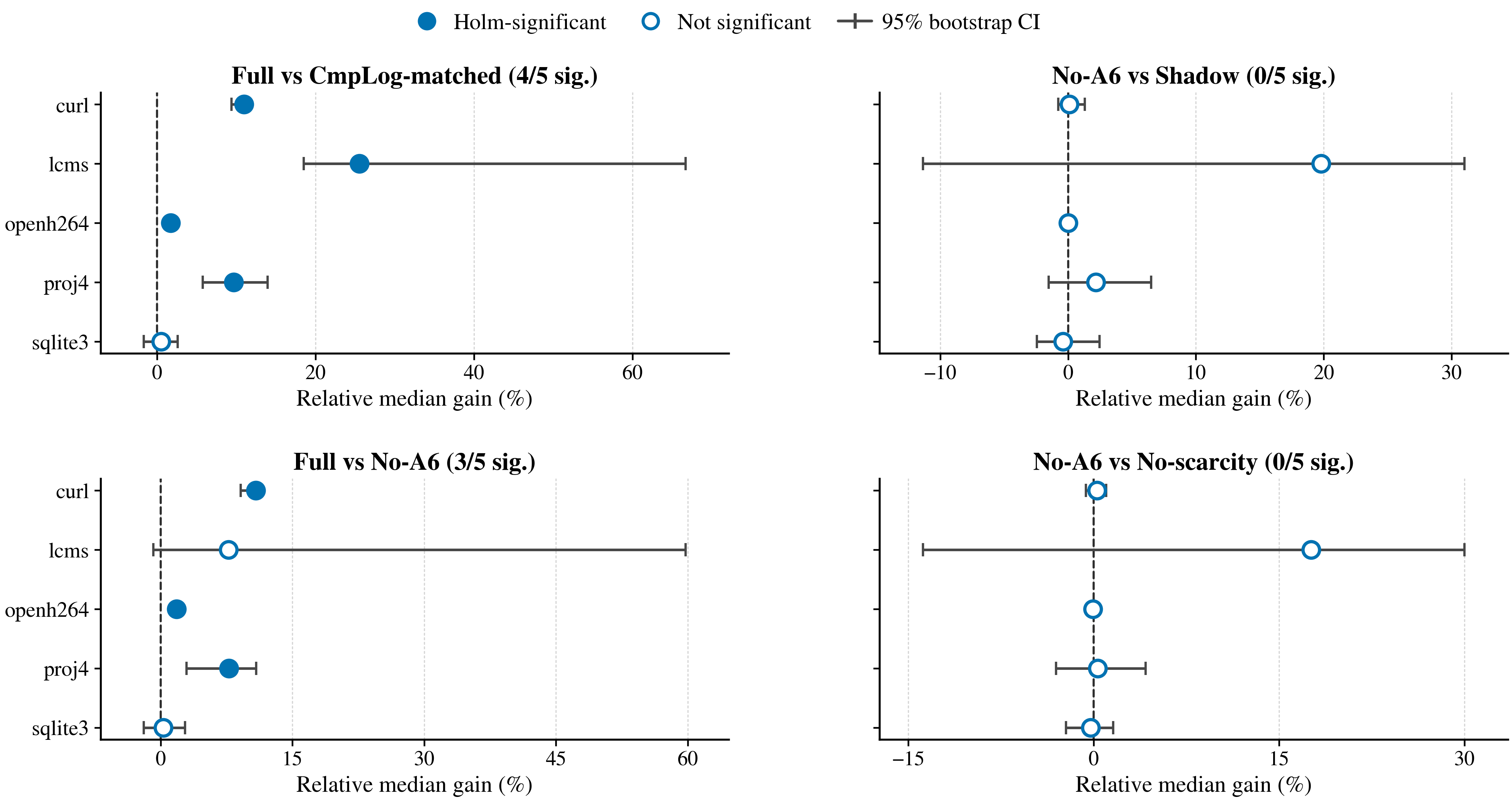}
\caption{Selected Batch B contrasts at 85,500 seconds. Points show relative median gain and horizontal bars show deterministic 95\% bootstrap confidence intervals. Filled markers indicate Holm-adjusted significance within the estimand-defined family; open markers are not significant. Intervals are independent percentile bootstrap summaries and are descriptive. Panel axes use separate scales to preserve small effects.}
\label{fig:batch-b}
\end{figure*}

Figure~\ref{fig:batch-b} shows the key estimand-defined contrasts.

\subsection{RQ5: Controller-Boundary Latency and End-to-End Rate Context}
Controller-compute medians are 317--824~$\mu$s; run-level compute P99 medians are 740--6491~$\mu$s, at most about 0.13\% of a 5,000~ms window. Per-target run-level quantiles and campaign-rate summaries are retained in the prepared artifact package.

Complete-boundary P99 medians, including synchronous logging, are below 14.7~ms, or 0.30\% of the window; audit-log-write P99 medians are below 13.9~ms. These measurements characterize only the five-second boundary path, not total \AdaRare overhead. Retained campaign-rate summaries vary from -27.99\% to +44.83\% because they combine \CmpLog, policy-induced paths, dictionary activity, queue decisions, energy shaping, and logging; fixed wall-clock coverage remains the relevant effectiveness metric.

\subsection{Unmatched Real-World Deployment Case Study}
Beyond \FuzzBench, an unmatched embedded-firmware deployment produced \AdaRare-generated inputs that exposed five distinct memory-corruption findings. Each was reproduced in a separate environment and later assigned a CVE identifier. Identifying details are omitted from this public preprint because they remain disclosure-sensitive. The cases provide evidence of practical applicability, but the deployment was not a matched vulnerability-yield experiment against \AFLpp or another fuzzer.

\section{Discussion and Limitations}
\noindent\textbf{Claim--evidence boundary.} Full \AdaRare has higher median edge coverage than a CmpLog-matched \AFLpp substrate on all five controlled follow-up targets, with four Holm-significant comparisons; this is the strongest controlled claim. Main is broader integrated-system evidence, and Batch A is diagnostic evidence for telemetry-guided selection within A1--A5. Batch B does not establish independent No-A6-versus-Shadow or scarcity-bundle effects; A6 evidence is target-dependent and weaker under global correction. We claim neither calibrated action values, superiority over unexecuted adaptive systems, nor comparative vulnerability yield.

Phases are analyzed separately; absolute edge levels are not compared across them. The five-target follow-up was chosen after Main without a preserved pre-outcome rationale, so it is not outcome-blind and remains subject to selection bias despite including three Main-significant and two Main-nonsignificant targets. Main uses 86,400 seconds; Batches A/B retain the protocol-recorded 85,500-second endpoint, with 86,400-second sensitivity.

The implementation borrows disjoint \LinUCB algebra, but action-induced telemetry and profile-conditioned targets make the scorer an arm-local ranking mechanism, not a calibrated contextual-bandit estimator. The mixed-domain normalizer, dormant escalation checks, unused legacy telemetry, and nonbinding caps specify the evaluated implementation; none is an independently validated mechanism.

Evidence levels differ: the packaged subset does not identify every historical runtime image; some variant semantics and override absences are author-confirmed, while run-linked wrappers, exact environments, and complete hardware records are missing (Appendix~\ref{app:evidence}). Parameters are fixed engineering choices without a complete retained pre-Main sensitivity record and are not claimed optimal.

With ten trials per cell, nonsignificance is not equivalence. Coverage is not a bug oracle. The unmatched firmware case study produced five separately reproduced, CVE-assigned findings but was not matched against other fuzzers, so neither study estimates comparative vulnerability yield.

\Needspace{5\baselineskip}
The remaining open limitations are direct contemporary adaptive baselines, a complete A6-by-scarcity factorial design, broader predeclared targets, later \AFLpp versions, and ground-truth bug benchmarks.

\section{Conclusion}
\AdaRare provides telemetry-guided joint closed-loop control of five \AFLpp surfaces. Full has higher median edge coverage than a CmpLog-matched substrate on all five follow-up targets; Main supplies broader integrated evidence and Batch A diagnostic evidence over simple schedules. Independent scarcity-bundle and No-A6-versus-Shadow effects are unestablished, and A6 evidence is target-dependent. An unmatched firmware deployment yielded five responsibly disclosed, CVE-assigned findings, supporting practical applicability but not comparative yield.
\section{Generative AI Use Disclosure}
OpenAI GPT-5.6 Thinking and GPT-5.6 Pro assisted with manuscript organization, language revision, the anonymized description of the real-world findings, and portions of the Python analysis/plotting code. The authors specified and independently verified all technical content, supplied the method, implementation, experimental design, and raw data, executed all experiments, and accept responsibility for the manuscript. No generative tool produced experimental observations, benchmark binaries, vulnerability evidence, or runtime evidence.

\section{Ethics Considerations}
The benchmark and firmware tests ran in isolated, locally controlled environments; no production service was targeted. For the real-world findings, we minimized and deduplicated triggering inputs, reproduced each behavior in a separate environment, inspected sanitizer output and stack traces, and used coordinated disclosure channels before public release. All five findings later received CVE identifiers. Identifiers, vendor/product details, endpoints, function names, advisory links, proof-of-concept inputs, and exploit steps are omitted from this public preprint because they remain disclosure-sensitive. The prepared artifact package excludes credentials, private server locations, raw crash inputs, and disclosure-sensitive materials.

\appendices
\section{Endpoint Sensitivity}
\label{app:endpoint}
All three exports contain the complete 86,400-second row for every campaign. Main uses 86,400 seconds as its primary endpoint. Batches A and B retain 85,500 seconds as their frozen inferential endpoint; their complete 86,400-second observations are reported only as a sensitivity analysis. Directions and within-family Holm-significance counts are unchanged.
\begin{table}[H]
\caption{Endpoint sensitivity. Entries are ``higher-median tests / Holm-significant tests.'' Main uses 86,400 seconds as its primary endpoint; Batches A and B use 85,500 seconds as their frozen primary endpoint, with the other column serving as the endpoint sensitivity.}
\label{tab:endpoint-sensitivity}
\centering\setlength{\tabcolsep}{3pt}
\begin{tabular}{lcc}
\toprule
Comparison family & 85,500 s & 86,400 s \\
\midrule
Main: AdaRare vs AFL++ & 8/8, 5/8 & 8/8, 5/8 \\
Batch A: controller & 15/15, 13/15 & 15/15, 13/15 \\
Batch A: A6 & 5/5, 2/5 & 5/5, 2/5 \\
Batch B: system fairness & 13/15, 8/15 & 13/15, 8/15 \\
Batch B: substrate diagnostics & 6/10, 2/10 & 6/10, 2/10 \\
Batch B: scarcity bundle & 3/5, 0/5 & 3/5, 0/5 \\
Batch B: A6 & 5/5, 3/5 & 5/5, 3/5 \\
\bottomrule
\end{tabular}
\end{table}

\FloatBarrier

\section{Secondary Time-Scale Checks}
\label{app:secondary}
The 6-hour, 12-hour, and normalized-AUC views support the same cautious interpretation. Main and Batch A profile-selection directions are already broad early in the run. Batch B continues to support Full against the matched controls and A6, while scarcity-bundle removal remains non-significant at every reported scale.
\begin{table*}[t]
\caption{Secondary time-scale checks. Entries are ``higher-median tests / Holm-significant tests.'' AUC is the trapezoidal area under edge coverage divided by the corresponding primary time budget (86,400 seconds for Main and 85,500 seconds for Batches A and B).}
\label{tab:secondary-endpoints}
\centering\setlength{\tabcolsep}{3pt}
\begin{tabular}{lcccc}
\toprule
Comparison family & 6 h & 12 h & Primary endpoint & Normalized AUC \\
\midrule
Main: AdaRare vs AFL++ & 8/8, 4/8 & 7/8, 4/8 & 8/8, 5/8 & 8/8, 5/8 \\
Batch A: controller & 15/15, 12/15 & 15/15, 12/15 & 15/15, 13/15 & 15/15, 12/15 \\
Batch A: A6 & 5/5, 1/5 & 4/5, 2/5 & 5/5, 2/5 & 5/5, 2/5 \\
Batch B: system fairness & 12/15, 8/15 & 14/15, 8/15 & 13/15, 8/15 & 13/15, 8/15 \\
Batch B: substrate & 5/10, 2/10 & 4/10, 3/10 & 6/10, 2/10 & 6/10, 3/10 \\
Batch B: scarcity bundle & 3/5, 0/5 & 3/5, 0/5 & 3/5, 0/5 & 4/5, 0/5 \\
Batch B: A6 & 5/5, 4/5 & 5/5, 4/5 & 5/5, 3/5 & 5/5, 4/5 \\
\bottomrule
\end{tabular}
\end{table*}

\FloatBarrier

\section{Multiplicity Sensitivity}
\label{app:holm}
The primary families are defined by estimand: Batch A has 15 controller, 25 static-profile, and five A6 contrasts; Batch B has 15 system-fairness, ten substrate-diagnostic, five scarcity-bundle, and five A6 contrasts. A global sensitivity deliberately treats all contrasts in a batch as one family. At 85,500 seconds, four Batch A contrasts lose significance under global correction because their adjusted $p$-values rise above 0.05: \texttt{proj4} No-A6 vs Constant, \texttt{lcms} No-A6 vs Random, \texttt{lcms} No-A6 vs Static A4, and \texttt{curl} Full vs No-A6. In Batch B, the adjusted $p$-value for \texttt{proj4} Full vs No-A6 likewise rises above 0.05. Complete raw, estimand-defined-Holm, and global-Holm values are retained in the prepared artifact package under \path{generated/}.
\begin{table}[H]
\caption{Multiplicity sensitivity at the controlled-batch primary endpoint of 85,500 seconds. Estimand-defined Holm families are reported; ``Global'' applies one correction to all 45 Batch A or 35 Batch B contrasts.}
\label{tab:holm-sensitivity}
\centering\setlength{\tabcolsep}{3.0pt}
\begin{tabular}{lcc}
\toprule
Contrast group & Estimand-defined & Global \\
\midrule
Batch A controller & 13/15 & 11/15 \\
Batch A static & 21/25 & 20/25 \\
Batch A A6 & 2/5 & 1/5 \\
Batch B fairness & 8/15 & 8/15 \\
Batch B substrate & 2/10 & 2/10 \\
Batch B scarcity bundle & 0/5 & 0/5 \\
Batch B A6 & 3/5 & 2/5 \\
\midrule
Batch A total & 36/45 & 32/45 \\
Batch B total & 13/35 & 12/35 \\
\bottomrule
\end{tabular}
\end{table}

\FloatBarrier

\section{Heterogeneous Scheduler Context}
\label{app:heterogeneous}
The six additional implementations provide heterogeneous scheduler context rather than a controlled SOTA ranking. Their per-target medians are reported below; full time series and pairwise statistics are retained in the prepared artifact package and are not used for a superiority claim.
\begin{table*}[t]
\caption{Median edge coverage at 86,400 seconds for the aligned seven-fuzzer main evaluation (ten trials per cell). Bold marks the highest median per target. These heterogeneous implementations provide scheduler context; only AdaRare versus AFL++ is treated as the primary integrated comparison.}
\label{tab:scheduler-baselines}
\centering\setlength{\tabcolsep}{2.4pt}
\begin{tabular}{lrrrrrrr}
\toprule
Target & AdaRare & AFL++ & AFLFast & EcoFuzz & Entropic & MOpt & T-Sched. \\
\midrule
\texttt{bloaty} & \textbf{6531.5} & 6367 & 5976.5 & 5757.5 & 5939 & 6124 & 4962.5 \\
\texttt{curl} & \textbf{11567} & 11208 & 9188.5 & 8973 & 8731 & 9596.5 & 10240 \\
\texttt{freetype2} & \textbf{11729.5} & 11552 & 7878 & 8148 & 10364.5 & 7985 & 11607.5 \\
\texttt{lcms} & \textbf{2347.5} & 1586 & 559.5 & 946 & 2058.5 & 847 & 2057 \\
\texttt{openh264} & \textbf{9828} & 9460.5 & 9556 & 9322 & 9559.5 & 9536.5 & 9575.5 \\
\texttt{proj4} & 7896.5 & 7612 & 734 & 774.5 & \textbf{7993} & 768.5 & 6424.5 \\
\texttt{sqlite3} & 19398 & 18472 & 18816 & 13084.5 & 17472 & 18969 & \textbf{19430.5} \\
\texttt{systemd} & \textbf{246} & 239 & 220 & 213.5 & 174 & 220 & 234.5 \\
\bottomrule
\end{tabular}
\end{table*}

\FloatBarrier

\section{Batch B Cell Medians and Average Ranks}
At 85,500 seconds the descriptive average ranks are Full 1.4, No-A6 3.0, No-scarcity 3.8, CmpLog-matched 3.8, Shadow 3.9, and default AFL++ 5.1. Ranks are computed within target and averaged only descriptively; absolute edge counts are never pooled across targets.
\begin{table*}[t]
\caption{Batch B matched-control medians at 85,500 seconds (ten trials per cell). Bold marks the highest median per target. All variants share byte-identical target, CmpLog, dictionary, and effective initial-corpus artifacts within each target.}
\label{tab:batch-b-medians}
\centering\setlength{\tabcolsep}{3.0pt}
\begin{tabular}{lrrrrrr}
\toprule
Target & Full & No-A6 & No-scarcity & CmpLog-matched & Shadow & AFL++ \\
\midrule
\texttt{curl} & \textbf{12557.5} & 11331 & 11300.5 & 11314 & 11318.5 & 11126 \\
\texttt{lcms} & \textbf{2031.5} & 1885.5 & 1603.5 & 1618 & 1574 & 1479.5 \\
\texttt{openh264} & \textbf{9730} & 9557.5 & 9563 & 9566 & 9557 & 9557 \\
\texttt{proj4} & \textbf{7755.5} & 7195.5 & 7172.5 & 7071.5 & 7041.5 & 5408 \\
\texttt{sqlite3} & 20234 & 20170.5 & 20219 & 20125.5 & \textbf{20251.5} & 20248 \\
\bottomrule
\end{tabular}
\end{table*}

\FloatBarrier

\section{Evaluated Controller Configuration}
\label{app:config}
\texttt{FORMAL-R1} denotes the frozen formal controller configuration that specifies the evaluated defaults. \texttt{RELEASE-R1} is the internal label for the packaged 15-file method-relevant source subset; it does not denote a public release. Neither label establishes byte identity with every historical runtime image. The Full campaigns used the defaults specified by \texttt{FORMAL-R1}. Apart from AdaRare activation, explicit variant switches, telemetry/audit controls, \CmpLog activation, and ordinary FuzzBench launch settings, the protocol did not override profile definitions, telemetry scales, controller-target parameters, the signal guard, P90 estimators, energy shaping, linear-scorer constants, or A6 constants. The profiles were not retuned after Main and no per-target parameterization was used. This is author-confirmed protocol evidence; the source-default structure is independently specified, but run-linked \texttt{exact\_env.tsv} files were not retained.

The queue-entry energy normalizer is separate from the per-effective-profile controller-target rate reservoirs. One global size-1024 auxiliary score reservoir receives queue-entry scarcity scores during energy assignment and one raw audit-target observation per completed window (named reward in the source/logs). It uses uniform reservoir sampling, requires 32 observations for a valid empirical P90, falls back to 1.0 otherwise, and resets every 256 windows. Thus the energy denominator mixes two internal score domains and is not interpreted as a calibrated queue-scarcity quantile.

Selection guards use controller-target and $\log(1+\text{edge-rate})$ EMAs with coefficient 0.20 and slope EMAs with coefficient 0.25. After at least eight selections, both absolute slopes below 0.002 and 0.005 mark a stagnating trend. New coverage/bits reset the stagnation counter; otherwise it rises under a stagnating trend and decays by one outside it. In the formal configuration, the effective revisit threshold is 24 windows throughout. The source contains nominal $72$- and $192$-window escalation checks keyed to discounted \texttt{total\_rounds}; however, initialized at $R_0=0$, with discount $0.999$ and one unit of aggregate pull credit per completed window, the state satisfies $R_{t+1}=0.999R_t+1$ and hence $R_t=1000(1-0.999^t)<1000$ for every finite $t$, so the 2000/10,000 checks are unreachable in the reported runs. After a 180-second cooldown, a threshold crossing revisits the non-current arm with the highest discounted mean controller target. During cooldown, stagnation can raise the exploration bonus to at most $1.35\times$. Slow-arm and repeated-zero-signal guards multiply candidate scores by bounded factors no lower than 0.80. The packaged snapshot source also updates a legacy \texttt{mix\_p} telemetry state, but the formal A6 top-$k$/softmax routing path does not consume it; it is therefore not listed as an evaluated routing mechanism.
\begin{table*}[t]
\caption{Evaluated controller configuration. Values are specified by \texttt{FORMAL-R1}; they are fixed engineering choices, not claimed optima, and no per-target controller parameters were used.}
\label{tab:evaluated-config}
\centering
\renewcommand{\arraystretch}{1.04}
\setlength{\tabcolsep}{3.0pt}
\begin{tabularx}{\textwidth}{@{}L{0.18\textwidth} L{0.43\textwidth} Y@{}}
\toprule
Subsystem & Evaluated values & Operational meaning \\
\midrule
Window/telemetry & 5000~ms; EMA $0.30$; scales $(0.2,0.5,0.2,1.0,2.0,1.0)$; feature cap $3.0$ & Six action-induced features, transformed and smoothed once per completed window. \\
Scarcity activity & path-length normalization $1/\max(|S|,1)$; campaign-lifetime $H_s$; no $H_s$ decay & Per-execution activity of nonzero slots, weighted by historical virgin-bucket discovery counts. \\
Queue scarcity score & observation EMA $1.0$; selection-time decay $0.995$; cap $10^9$ & Queue-entry state consumed by bounded energy shaping. \\
Controller-target rate scales & per-effective-profile uniform reservoirs, size 1024; post-warmup and 32 positive samples; edge/scarcity/CmpLog fallbacks $50/5/10$ & Running P90 engineering scales; a current positive observation is inserted before that window's P90 is evaluated; not rolling, decayed, or phase-local. \\
Auxiliary energy normalizer & one global size-1024 reservoir; minimum 32; fallback $1.0$; reset every 256 windows & Receives queue scarcity observations and one raw audit-target observation/window (source/log label: reward); its P90 is a mixed-domain engineering normalizer, not a calibrated scarcity quantile. \\
Recency-weighted linear scorer & $\alpha=0.6$; ridge $10$; discount $0.999$; exploration-term cap $2$; score cap $5$ & Disjoint arm-local models with finite-value checks; matrix cap $10^{12}$ and rescale factor $10^{-6}$. \\
Controller-target bounds & raw interval $[-0.25,1.25]$; model target $[0,1]$; zero-signal term $0.015$; cost factor $0.10$ when $I_t^{\mathrm{sig}}=1$ & Negative raw audit values are floored to zero before the arm-local model update. \\
Throughput reference & update weight $0.1$ when throughput rises or during warmup; $0.001$ when it falls & Asymmetric EMA provides a slowly decreasing cost reference. \\
Execution-time gate & $\eta_g=1$; $g_{\min}=0.05$; execution-time EMA current-sample weight $0.01$; bonus multiplier $0.02$; hard factor cap $1.15$; positive threshold $10^{-4}$ & The first measured execution initializes the baseline; later executions update it before gate evaluation as $B_i=0.99B_{i-1}+0.01\,\mathrm{exec}_i$. Row normalization gives $r_t^+\leq25/27$ and $h_t/r_t^+\geq0.35$ when active, while $b_t\leq0.02\log 2\approx0.0139$; neither cap binds. \\
Profile energy modes & A1 $1+0.5z$ (cap 2); A2 $1+2\log(1+2z)$ (cap 5); A3 $1+z$ (cap 3); A4 $1+2(z-0.8)_+$ (cap 3); A5 $1$ & $z$ divides the queue score by the auxiliary-reservoir P90; A6 inherits the delegated base mode. \\
Queue/mutation bounds & dictionary 0--100; havoc factor $0.95$--$1.20$ per scaling site; final \texttt{perf\_score} cap $100\!\times\!\texttt{HAVOC\_MAX\_MULT}=6400$; queue rejection cap 8 & The packaged source consumes the same profile havoc factor at two sequential score-scaling sites; Table~\ref{tab:arms} reports the per-site value. Favored/new predicates are conjunctive during retries; after eight failures, the current valid candidate is accepted. \\
Warmup/dwell & two pulls/arm; dwell threshold one window; zero-signal emergency streak 3 & Dwell can be bypassed by warmup, forced revisit, active trend, or emergency paths. \\
Trend/revisit guards & controller-target/edge EMA $0.20$; slope EMA $0.25$; minimum 8; slope thresholds $0.002/0.005$; effective formal stagnation threshold 24; cooldown 180~s & Nominal 72/192 escalation checks are unreachable under the formal discounted round state; eligible stagnation revisits the non-current arm with the highest discounted mean controller target. \\
Score guards & stagnation bonus cap $1.35$; slow ratio $0.70$; repeated slow/zero-signal score-factor floor $0.80$ & Guardrails multiply exploration bonuses or candidate scores; they do not change the external metric. \\
A6 routing & top-$K=2$; softmax $T=1.20$; for $j\in\mathcal{K}_t$, $\tilde p_j=\max(p_j,0.05/|\mathcal{K}_t|)$, then $p'_j=\tilde p_j/\sum_{\ell\in\mathcal{K}_t}\tilde p_\ell$ & $\mathcal{K}_t$ is the current top-$K$ base-arm set ranked by discounted mean controller target; the probability floor is applied before renormalization, and lower arm index resolves exact mean ties. \\
A6 delegated sharing & $w=\min(1/\max(\pi,10^{-6}),2)$ & Selected A6 receives the ordinary selected-meta-arm update; only the delegated effective profile receives clipped additional credit using the same controller target. \\
\CmpLog & producer after real \CmpLog child; minimum gain $0.25$; window clip $8$; fallback scale $10$ & Integer-distance/equal-byte improvements only; ordinary coverage and queue growth are separate. \\
\bottomrule
\end{tabularx}
\end{table*}

\FloatBarrier

\section{Evidence Status}
\label{app:evidence}
The evidence table prevents source-level specification, archived runtime evidence, and author confirmation from being conflated. In particular, the packaged source subset labeled \texttt{RELEASE-R1} does not by itself prove the identity of every historical runtime image.
\begin{table*}[t]
\caption{Evidence status for central reproducibility and attribution claims. ``Author-confirmed'' is intentionally separated from independently archived run linkage.}
\label{tab:evidence-status}
\centering
\renewcommand{\arraystretch}{1.04}
\setlength{\tabcolsep}{3.2pt}
\begin{tabularx}{\textwidth}{@{}L{0.27\textwidth} L{0.32\textwidth} Y@{}}
\toprule
Claim & Preserved evidence & Limitation \\
\midrule
Export completeness and reported statistics & packaged projections, hashes, deterministic scripts, complete cells/snapshots & raw cloud storage metadata is redacted or not retained \\
Batch B target/CmpLog/dictionary/effective-corpus identity & artifact manifest and SHA-256 equality within target & original variant wrapper and runtime-image digest are absent \\
CmpLog-matched \texttt{-c} activation & archived runtime argument/configuration evidence and author confirmation & complete image-to-source mapping is absent \\
Shadow computes but does not apply profiles & author-confirmed frozen semantics and smoke/runtime counters (candidate $>0$, applied $=0$) & run-linked wrapper source is not retained \\
No-A6 and No-scarcity semantics (runtime label No-rarity) & author-confirmed frozen definitions and smoke/runtime feature gates & the preserved evidence does not independently reconstruct every disabled path \\
Batch A Constant/Random/Round-robin/Static semantics & frozen Batch A fixed-context definition plus source/config policy semantics and complete-cell data & original run-linked wrapper/image mapping is incomplete; shared-substrate strength is not upgraded beyond preserved evidence \\
Formal controller parameters and no post-Main tuning & source-default structure plus author-confirmed protocol; no per-target values & run-linked \texttt{exact\_env.tsv} and pre-run override manifest are not retained \\
Prepared artifact source and navigation subset & complete E04 AFL++ source tree plus 15 method-navigation files and hashes in \texttt{source\_release/RELEASE-R1} & neither source package establishes identity with every historical runtime image \\
Hardware allocation & author-retained cloud allocation records & OS/kernel, pinning, topology, and worker allocation are not machine-verified \\
Boundary-latency design & redacted matched-environment manifest and independent analysis & trials were not paired by seed, host slot, or pair identifier \\
\bottomrule
\end{tabularx}
\end{table*}

\FloatBarrier

\section{Standard Data, Source, and Reproduction}
\label{app:reproduction}
A separate prepared artifact package includes anonymized projections of the three standard exports under \path{data/}, deterministic analysis and plotting code under \path{analysis/}, complete statistical outputs under \path{generated/}, and the TeX tables and plots used by the manuscript. The projections remove private or analysis-irrelevant metadata while preserving revision label, fuzzer, benchmark, trial, snapshot time, edge coverage, crash key, and bug-count fields. Crash-key and bug-count columns are retained for completeness but are not analyzed because they do not constitute a validated ground-truth vulnerability oracle. Their SHA-256 values are:
\begin{itemize}
\item Main 800-campaign projection: \texttt{\seqsplit{feebf5f8cbdbd481e88bd7ec56df3149948d7a2b1d52067eb2c0599e664e8eb2}}.
\item Batch A 500-campaign projection: \texttt{\seqsplit{836057133f9dd8a50044716da6026a6feb3a6d9b6b0aaea78042b57d82928778}}.
\item Batch B 300-campaign projection: \texttt{\seqsplit{8758649cb552e576356225f8d16bf675f33a7de816c3ad7619520a24ba687c2f}}.
\end{itemize}
Each has 97 snapshots per campaign from 0 through 86,400 seconds, ten trials per cell, no duplicate benchmark--variant--trial--time rows, and monotonically non-decreasing edge coverage. The standard raw Batch A export has SHA-256 \texttt{\seqsplit{7a3920bd49e75bad640ba0fccb7ea2426652012c60be91c77fca42f7587424bc}}; the packaged Batch A file is its anonymized analysis projection.

The prepared artifact package contains the complete E04 \AFLpp source tree. For method navigation, we also provide the packaged 15-file subset labeled \texttt{RELEASE-R1} under \path{source_release/}, together with a source-level specification and file hashes. The integrated controller semantics were maintained across sequential phases, while intervening commits added controls and FuzzBench/environment wiring. Complete runtime-image-to-source cryptographic mappings were not retained, so the snapshot is not presented as proof of every historical executable image. Parameter and variant author confirmations, along with their independent-verification limits, are preserved under \path{provenance/}.

\end{document}